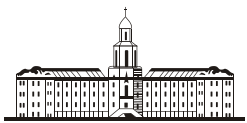

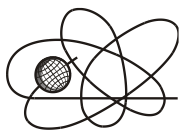

РОССИЙСКАЯ АКАДЕМИЯ НАУК

**ИНСТИТУТ ПРОБЛЕМ БЕЗОПАСНОГО РАЗВИТИЯ АТОМНОЙ ЭНЕРГЕТИКИ**

RUSSIAN ACADEMY OF SCIENCES

**NUCLEAR SAFETY INSTITUTE**

Препринт ИБРАЭ № IBRAE-2018-13

Preprint IBRAE-2018-13

**Arutyunyan R.V., Vasiliev A.D.**

# TRANSITIONS BETWEEN VOLUME-LOCALIZED ELECTRON QUANTUM LEVELS OF FULLERENE C$_{60}$ ION

Москва
2018

Moscow
2018

# Transitions between Volume-Localized Electron Quantum Levels of Fullerene $C_{60}$ Ion

Rafael V. Arutyunyan[1] and Alexander D. Vasiliev[1]

[1] Nuclear Safety Institute, B.Tulskaya 52, Moscow 115191, Russia.

Correspondence should be addressed to Alexander D. Vasiliev; vasil@ibrae.ac.ru

**Abstract**

The excited short-lived volume-localized electron quantum levels (VLELs) existent due to Coulomb potential well inside positive ion $C_{60}^{+Z}$ are analytically investigated in the paper using a simplified spherical fullerene model. Hence, those electron levels appear exclusively after the ionization of neutral fullerene taking into account the unique geometrical shape of sphere. The existence of those levels is argued, and their basic parameters (the energy levels, the wave eigenfunctions) are approximately calculated. The wave functions of VLELs are basically localized inside fullerene ion sphere (with a maximum amplitude in the centre) in contrast to ordinary surface-localized electron levels (SLELs) having a wave functions in the vicinity of fullerene sphere formed by the cluster composed of carbon ions. Contrary to VLELs, the wave functions of SLELs are present both in charged and neutral fullerene.

The analysis of electron beam interaction with the medium consistent of fullerenes ions $C_{60}^{+Z}$ is conducted as the application of the methods developed. The analytical dependencies of free electron recombination cross-sections for the capture to the volume-localized electron levels are obtained. It is shown that the probabilities of electron capture to these VLELs are considerably larger compared to capture to SLELs.

Also, the calculational results of dipole moments for quantum transitions from fullerene ions VLELs to other VLELs and to SLELs with spontaneous photon emission are also presented in the paper. The calculated dipole moments depend on fullerene ionization extent, initial and final electron states, and are varied from about 0.2 to 5 in atomic system of units. Finally, the principal possibility of coherent radiation generation on fullerene ions' VLELs is discussed.

Thus, the electrons captured on these discrete levels of fullerene form a sort of a short-lived "nano-atom" or "nano-ion", in which the electrons are localized inside a positively charged spherical "nucleus".

**Introduction**

New allotropic forms of carbon such as fullerenes, fullerites, onion-like fullerenes and afterward carbon nanotubes, graphene, doped and endohedral fullerenes, have been discovered respectively recently [1]. Currently, those nanomaterials are considered as perspective materials in different areas of technology. The review of fullerenes' physical properties and behaviour including fullerene ions under collisions with fullerenes and other particles is presented in the paper [2].

The ionization cross-sections during electron impact [3,4] or photoionization cross-sections [5] of neutral and positively or negatively charged fullerenes are obtained in the literature. It should be noted that due to big size of fullerene compared to ordinary atoms and molecules and due to large number of carbon atoms in fullerene, the direct quantum-mechanical calculations of these objects are extremely complicated and have a low accuracy and poor prediction ability. This is why some simplified model assumptions are necessary for the analysis of physical properties of fullerenes and fullerene ions.

A number of investigators [6,7] successfully apply the so called "jellium" model to qualitatively describe the physical parameters from experimental observations. The authors calculated the energy levels spectrum for $C_{20}$ as well as for $C_{60}$ molecules. For example, the lowest occupied electron energy level corresponds to about -44 eV while the highest occupied level equals approximately to -4 eV in $C_{20}$ neutral fullerene. In $C_{60}$ neutral fullerene these levels are almost the same: -43 eV and -3.5eV respectively. In the paper [7] the potential of the well and the wave functions for fullerene $C_{60}$ are calculated. A spherical electron gas model is applied as a simple model of $C_{60}$ molecule to the optical scattering of electromagnetic wave on fullerene molecules [8].

The volume-localized electron levels (VLELs) existent due to Coulomb potential well inside positive $C_{60}^{+Z}$ ion have been reported for the first time in [9]. To our knowledge, these electron levels located primarily inside the fullerene ion spheroid have not clearly mentioned in the literature yet.

The system fullerene cation + electron and maybe neutral fullerene + electron can form the quantum coupled system similar to inverted nanoion ($C_{60}^{+Z} + e^-$) or nanoatom ($C_{60}^{+} + e^-$). The electron is localized at discrete energy levels inside the charged sphere of fullerene cluster composed of carbon ions. This is due to unique geometrical form of fullerene cluster as a spheroid resulting in Coulomb potential well formation inside the fullerene. It could be said, that due to this topology, the quantum VLELs are concentrated on the "wrong" side of fullerene sphere.

At the same time, the ordinary surface-localized electron levels (SLELs) are typical for charged and neutral fullerene. Thus, concerning the impact of the electron beam with fullerene ions gas, the capture of free electrons is possible both to VLELs and SLELs energy levels. Also, the spontaneous transitions from VLELs to other VLELs and to SLELs could take place with simultaneous light emission.

In the paper [10] the authors conduct the calculations of fullerenes recharge cross-sections within the bounds of standard recharge theory with tidy two-well symmetric potential. Those calculations are in a good agreement with experimental data.

There is direct experimental evidence that fullerene ion degree (the number of positive elementary charges on a fullerene) may reach the quantity of about +10. For example, the experiments are considered in [11], where high-charged fullerene ions were formed during ionization by electron beam. The ions were detected with positive charge corresponding up to 6 elementary charges per particle. In the paper [12] the experiments are reported, in which the stable 12-fold fullerene ions were observed (12 elementary charges per particle) after irradiation of fullerene jet by strong infrared laser impulse.

The fullerene ions are stable or meta-stable. For example, the authors of the paper [13] estimate numerically the characteristic $C_{60}^{+Z}$ ion lifetime to be of the order of several seconds for Z<+11. However, according to this investigation, the dramatic lowering of fullerene ions lifetime by 10 orders of magnitude takes place when increasing Z from +11 to +13. The ion lifetime longer about 0.5 μs is reported in the experimental work [11].

At the same time, the results of quantum-mechanical calculations by density functional theory (DFT) [14,15,16] show that the meta-stable fullerene ions can be obtained with ionization degree up to Z=+10.

The study of optical and nonlinear optical properties of fullerenes and fullerene ions with different ionization degree $Z$ including high $Z$ is very important because those properties have not been fully analyzed up to now. Currently, this topic attracts many investigators throughout the world, see, for example, [17]. The fullerenes, onion-like fullerenes and carbon nanotubes (CNT) are very perspective materials, particularly, due to their unique geometry. The discovery of new unexpected phenomena of those nanomaterials is still anticipated in the future investigations.

In this paper the simplified spherical model is used for qualitative and quantitative description of volume-localized electron levels existent due to Coulomb potential well inside positive $C_{60}^{+Z}$ ion.

The VLELs wave eigenfunctions and the energy levels are investigated in the next Section. Then, in the Section devoted to the results of investigation, the electron capture cross-sections during recombination are calculated on the basis of standard quantum-mechanical methods. Also, the main results of calculations of dipole moments for transitions from fullerene ions' VLELs to other VLELs and to SLELs are presented. In the last subsection, the principal possibility of coherent radiation generation on fullerene ions VLELs is discussed.

**Materials and Methods**

We use for the analysis the simplified spherical model of fullerene. According to this model, the total positive charge of carbon ions of fullerene cluster and total negative charge of bounded and delocalized electrons are uniformly distributed on spherical surface of infinitesimal thickness. So, the quantum-mechanical problem allows the axial-symmetrical formulation. This assumption considerably simplifies the consideration of volume-localized electron levels in fullerene ion.

**Volume-localized Electron Levels of Fullerene Ion**

Let us find the electron wave functions corresponding to different energy levels in the potential produced by charged fullerene. We consider the ion $C_{60}^{+Z}$ in the analysis.

We suggest that except ordinary electron level states bounded with carbon ions (localized electron quantum levels) and states unified with a whole cluster (delocalized electron quantum levels) characteristic of neutral fullerene there exist another delocalized electron levels in fullerene ion due to the effective Coulomb field of charged fullerene.

Indeed, within the bounds of this model the dependency of electron potential energy in a Coulomb potential of charged fullerene $U(r)$ is presented in Figure 1 as an example. Here $r$ is the distance to the centre of fullerene, $Z$ – the charge of fullerene, $e$ – the charge of electron, $\varepsilon_0$ – the dielectric permeability of vacuum, $r_\mathrm{f}$ – the radius of fullerene.

The depth of potential well $U_0$ is (see Figure 1):

$$U_0 = \frac{Ze^2}{4\pi\varepsilon_0 r_f}. \tag{1}$$

By inserting the following values $Z$=+10, $e \approx 1.60 \cdot 10^{-19}$ C, $\varepsilon_0 \approx 8.85 \cdot 10^{-12}$ C/(V·m), $r_f \approx 3.51 \cdot 10^{-10}$ m into this formulae, we get $U_0 \approx 6.57 \cdot 10^{-18}$ J $\approx 41.1$ eV. As long as the depth of the well is deep enough in comparison to the value

$$\frac{\hbar^2}{m_e(2r_f)^2} \approx 5\cdot 10^{-20}\,\text{J} \approx 0.31\text{ eV},$$

where $\hbar \approx 1.05\cdot 10^{-34}$ kg·m²/s – the reduced Planck constant, $m_e \approx 0.91\cdot 10^{-30}$ kg – the mass of the electron, $2r_f$ – the characteristic size of the well, then in this potential well a number of electron states exists which is much more than unity [18]. Note, that this conclusion is valid for every fullerene ion's charge from Z=+1 to Z=+10.

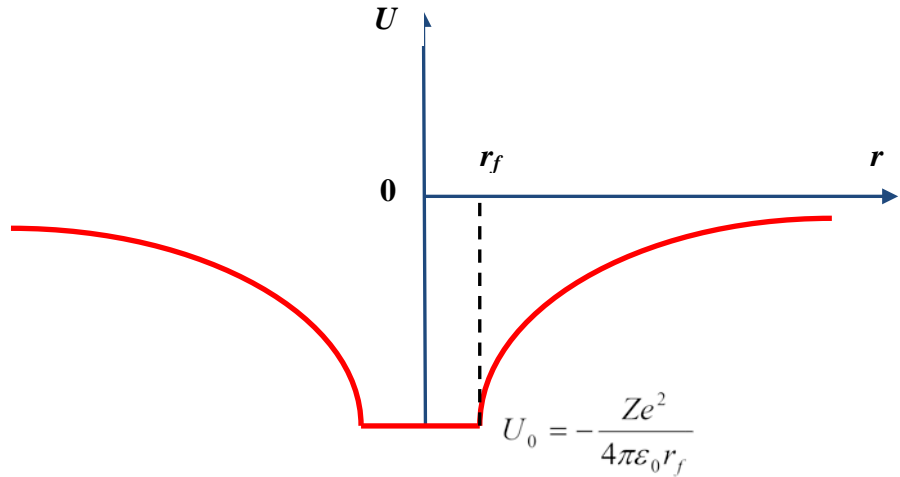

Figure 1: Simplified radial dependence of electron potential energy in Coulomb field of charged fullerene $C_{60}^{+Z}$.

Despite the "jellium" or simplified spherical model within which VLELs can be obtained is rather rough approximation, the existence of VLELs is not argued. In reality the resultant Coulomb field in fullerene ion is not spherically symmetric, it is modulated on angles $\theta$ и $\varphi$ (see Figure 2) correspondent to carbon ions locations in fullerene shell. But taking into account of this circumstance will just result in correction of wave states eigenvalues and spatial dependencies, while will not call in question the conclusion on the existence of VLELs itself.

For the axial symmetry case the solution for wave functions is presented as a product of radial and spherical functions:

$$\psi_{nlm} = R_{nl}(r)Y_{lm}(\theta,\varphi), \qquad (2)$$

where $n$, $l$ and $m$ are the principal, azimuthal and magnetic quantum numbers respectively. The number $n$ satisfies the non-equality $n \geq l+1$. The energy of electron $E_{nl}$ corresponds to every wave function.

The equations for radial wave function within the bounds of simplified spherical model of spherical fullerene in atomic system of units are written as

$$\frac{d^2 R_{nl}}{dr^2} + \frac{2}{r}\frac{dR_{nl}}{dr} - \frac{l(l+1)}{r^2}R_{nl} + 2\left(E_{nl} + \frac{Z}{r_f}\right)R_{nl} = 0, \qquad 0 \leq r < \frac{r_f m_e \alpha}{\hbar^2} = r_{full}, \qquad (3)$$

$$\frac{d^2 R_{nl}}{dr^2} + \frac{2}{r}\frac{dR_{nl}}{dr} - \frac{l(l+1)}{r^2}R_{nl} + 2\left(E_{nl} + \frac{Z}{r}\right)R_{nl} = 0, \qquad r \geq \frac{r_f m_e \alpha}{\hbar^2} = r_{full}. \tag{4}$$

where $\alpha = e^2/(4\pi\varepsilon_0)$, $Z$ – the fullerene ion charge, $r_{full} \approx 6.63$ – the fullerene radius in atomic system of units. We consider now the region inside the charged fullerene $C_{60}^{+Z}$ that is $r < r_{full}$.

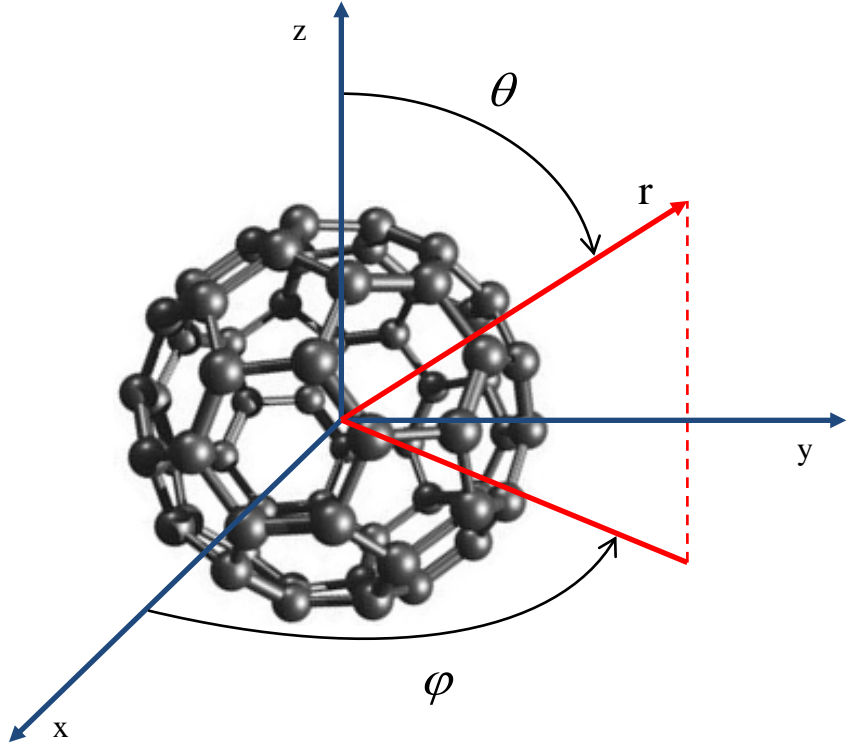

Figure 2: Fullerene ion $C_{60}^{+Z}$ in a spherical system of coordinates.

The Equation 3 can be rewritten as

$$\frac{d}{dr}\left(r^2 \frac{dR_{nl}}{dr}\right) = -\gamma_{nl} r^2 R_{nl} + l(l+1)R_{nl}, \qquad \gamma_n = 2\left(E_{nl} + \frac{Z}{r_f}\right). \tag{5}$$

In beginning, let us investigate electrons-states with $l=0$ (the spherical symmetry s-states). The solutions of (3) limited at $r=0$ will be

$$R_{n0} = \frac{\sin(\sqrt{\gamma_n}\, r)}{r} = \frac{\sin(k_n r)}{r}, \quad k_n = \sqrt{\gamma_n}. \tag{6}$$

Apart from (6), another solutions tending to infinity at $r=0$ will also be applicable:

$$R_{n0} = \frac{\cos(k_n r)}{r}. \qquad (7)$$

However, those solutions are physical too, because the integral of square of wave function module in the vicinity of singularity point $r=0$ is convergent, that is finite. Thus, the wave eigenfunctions inside fullerene sphere for azimuthal quantum number $l=0$ are as following:

$$R_{n0} = \frac{\sin(k_n r)}{r} + \eta_n \frac{\cos(k_n r)}{r} \qquad (8)$$

with correspondent energy levels $E_{n0} = -Z^2/(2\varsigma_n^2)$, $\varsigma_n = n = 1,2,3...$; the parameter $\eta_n$ is determined from boundary conditions.

The solutions of Equations 4 for the region outside the fullerene shell are presented in Appendix 1. The conditions of sewing together the wave functions and their derivatives inside and outside a fullerene sphere result in that for charge $Z=+1$ and $\varsigma_n \gg 1$ the parameter $\eta_n$ is equal about -0.24. The expression $E_n = -Z/(2\varsigma_n^2)$, where $\zeta_n$ is the integer number, arises from the conditions of converging of the solution at $r \to \infty$ and is the obligatory property of Coulomb systems.

The parameter $\gamma_n = 2(E_n + Z/r_{full})$ from Equation 5 should be more than zero otherwise $k_n = \sqrt{\gamma_n}$ will be imaginary. In last case the solution would be

$$R_{n0} = \frac{sh(\sqrt{|\gamma_n|}r)}{r} + \eta_n \frac{ch(\sqrt{|\gamma_n|}r)}{r} \qquad (9)$$

However, the appropriate energy levels in this case are lower than the bottom of Coulomb potential well (see Figure 1) and, consequently, the solutions (8) do not have a physical meaning.

The complete set of solutions for Equations 3 and 4 for the case $l>0$ and for the region outside the fullerene sphere are given in Appendix 1.

**Results and Discussion**

Let us now use the model of VLELs to some interesting quantum mechanical applications.

**Electron recombination with electron capture to volume-localized levels of fullerene ion $C_{60}^{+Z}$**

Consider the electron beam incident on the medium consisting of fullerene $C_{60}^{+Z}$ ions.

The intensity of electron beam passing through the medium containing particles is changing according to Bouger-Beer-Lambert law:

$$I_n(E,x) = I_0(E) \exp[-nQ(E)x], \qquad (10)$$

where $I_0$ is the initial intensity of electron beam, $1/(m^2 s)$; $n$ – the concentration of particles, $1/m^3$; $x$ – the distance, passed by the beam, m; $Q(E)$ - the total cross-section of electron scattering on the particle; $E$ – the energy of the electron.

Suppose that the medium contains fullerenes, fullerene ions or onion fullerenes. The electrons scatter on nanoparticles by two ways:
- Elastics cattering (when electron energy is not changing during collision but the electron can scatter at some angle from initial direction of propagation);
- Non-elastic scattering with energy loss of incident electron including several possible channels (fullerene transition to the state with higher electron energy, the scattering with excitation of fullerene's plazmon oscillations, one- or many-fold ionization of fullerene, the electron capture with transition to exciting state or photon emiting (electron recombination).

We will stay in more detail on electron capture phenomenon. Consider the problem of moving electron capture by positive ion of $C_{60}^{+Z}$ fullerene in the state of rest (Figure 3).

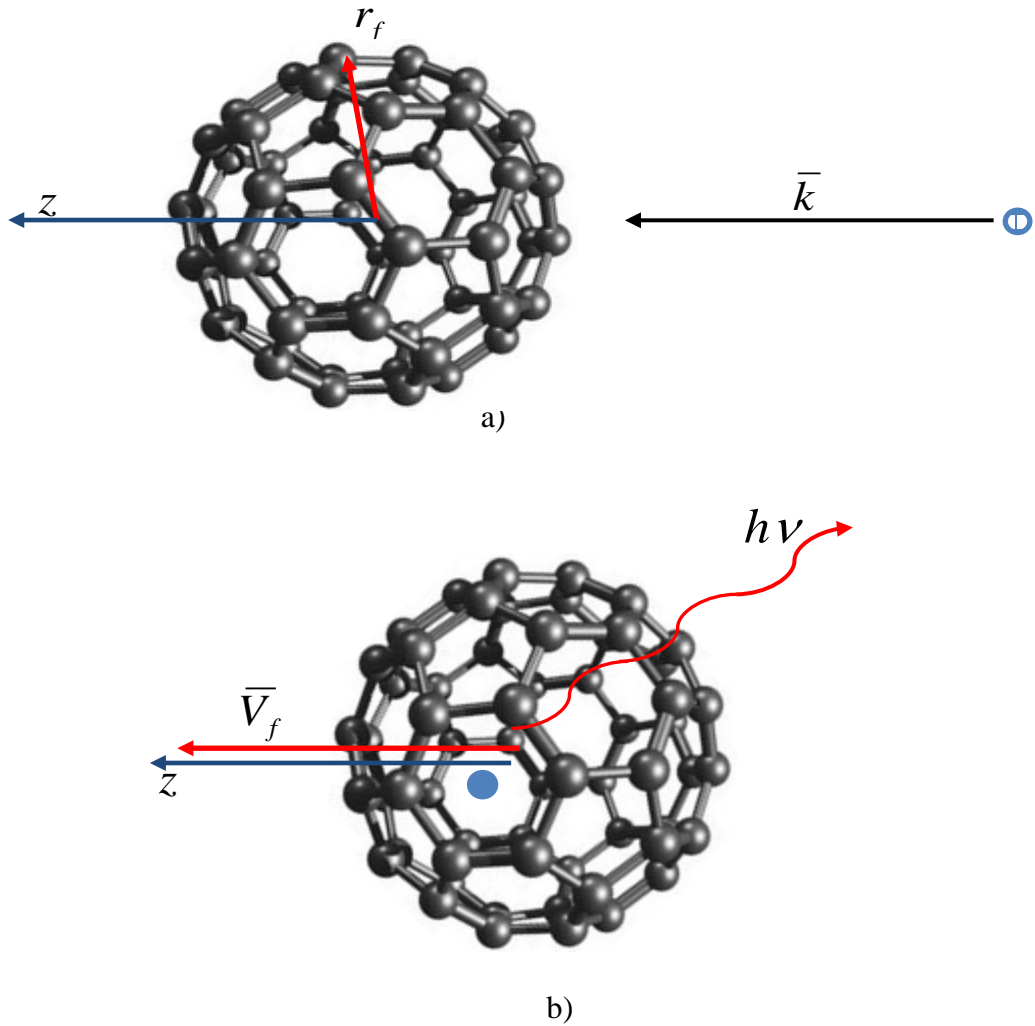

Figure 3: The geometry of electron recombination phenomenon: a) particles position before capture; b) particles position after capture.

The electron with wave vector $\bar{k}$ and momentum $\bar{p}$ directed along $z$-axis and kinetic energy $E = p^2/2m_e = k^2 \hbar^2 / 2m_e$ (Figure 3a) is falling upon the fullerene. As a result of interaction the electron can be captured by the fullerene that is the electron can transfer to one of unoccupied electron levels of neutral or charged fullerene with instantaneous emitting of photon (Figure 3b).

If we know the wave functions $\psi_n(\bar{r})$ corresponding to different energy quantum levels of fullerene ions, then free electron capture amplitude during transition from continuum spectrum state with energy $E$ to given energy level $E_n$ with wave function $\Psi_n$ is written by the following manner:

$$T(\bar{k},n) = \langle \psi_n(\bar{r}) | U(\bar{r}) | \exp(i\bar{k} \cdot \bar{r}) \rangle, \qquad (11)$$

where $U(\bar{r})$ is the potential of ion $C_{60}^{+10}$, presented in Figure 4, $\bar{k}$ - the wave vector of incident electron. Thus, we use the classical Born approximation. The potential of neutral fullerene [6] is shown in Figure 5.

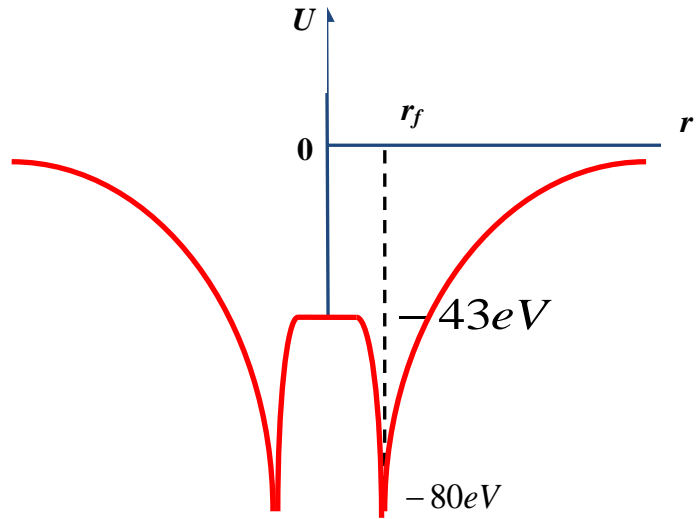

Figure 4: Schematic presentation of self-consistent potential of fullerene ion $C_{60}^{+10}$.

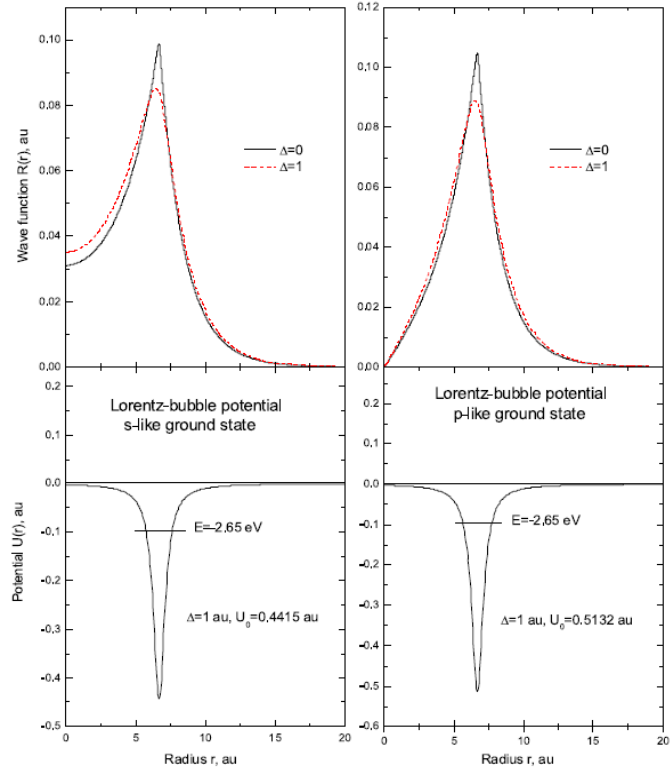

Figure 5: The wave functions and the potential of the well for $C_{60}$ shell reproduced from [6].

So, the amplitude of the electron capture on free electron levels of fullerene ion with wave function $\psi_n$ will be the integral on volume:

$$T(\bar{k}, n) = \oint_V \exp(ikz) U(\bar{r}) \psi_n(\bar{r}) dV, \qquad (12)$$

where $\bar{k}$ is the electron wave vector.

For estimation of this integral it is convenient to expand the plane wave of incident electron on spherical waves the following way [18] (see Figure 3a):

$$e^{i\bar{k}\cdot\bar{r}} = 4\pi \frac{1}{2k} \sum_{l=0}^{\infty} \sum_{m=-l}^{l} i^l R_{nl}(r) Y_{lm}^*\left(\frac{\bar{k}}{k}\right) Y_{lm}\left(\frac{\bar{r}}{r}\right), \qquad (13)$$

$$R_{nl}(r) = (-1)^l 2 \frac{r^l}{k^l} \left(\frac{d}{rdr}\right)^l \frac{\sin kr}{r}.$$

But the product of radial functions $R_{nl}(r)$ and spherical functions $Y_{lm}(\theta, \varphi)$ are exactly the wave eigenfunctions inside a charged fullerene sphere. This sircumstance considerably simplifies the approximate calculation of integrals (12).

The capture probability is the square of matrix element. As a result the square root from probability of electron capture to sphericall symmetrical level ($n=0$) is estimated as

$$\sqrt{P_{capture}} = \frac{4\pi}{16\pi^2} \cdot \frac{1}{2k} \cdot \frac{Ze^2}{4\pi\varepsilon_0 r_f} \cdot \frac{\hbar^2(4\pi\varepsilon_0)^2}{m_e Z^2 e^4} = \frac{1}{4\pi} \cdot \frac{1}{2k} \cdot \frac{\hbar^2 4\pi\varepsilon_0}{r_f m_e Z e^2}. \qquad (14)$$

The term $16\pi^2$ has arisen due normalization of spherical functions (see Appendix 2). In this expression the wave vector $k$ is expressed in atomic unite, so it is necessary to make the substitution

$$k \to k \cdot \frac{\hbar^2 4\pi\varepsilon_0}{m_e Z e^2}. \tag{15}$$

Then we get the simple expression

$$\sqrt{P_{capture}} = \frac{1}{16\pi^2} \frac{2\pi}{kr_f}. \tag{16}$$

From the above expression we concludethatthe capture cross-section (proportional to capture probability) is inversely proportional to square of wave vector or to electron kinetic energy.

The partial capture cross-section is

$$\sigma_{capture,n} = \frac{4\pi^2 \omega_n}{c} \left| T(\bar{k},n) \right|^2, \tag{17}$$

where $\omega$ is the angular frequency of emiting photon, $c$ – the light speed.

For more thourough calculation instead of (16) we need to take the following integral

$$\sqrt{P_{capture}} = \frac{1}{8\pi k r_f^2} \int_0^{r_f} \sin kx \cdot R_{nl}(r) Y_{lm}(\theta,\varphi) \cdot dx, \tag{18}$$

where $k$ is the wave vector of incident electron and the wave vector of electron level is written as

$$k_n = \sqrt{2\left(E_n + \frac{Z}{r_f}\right)}. \tag{19}$$

The total capture cross-section is

$$\sigma_{capture,total} = \sum_n \sigma_{capture,n}$$

After calculation we get for capture cross-section the following graph (Figure 6).

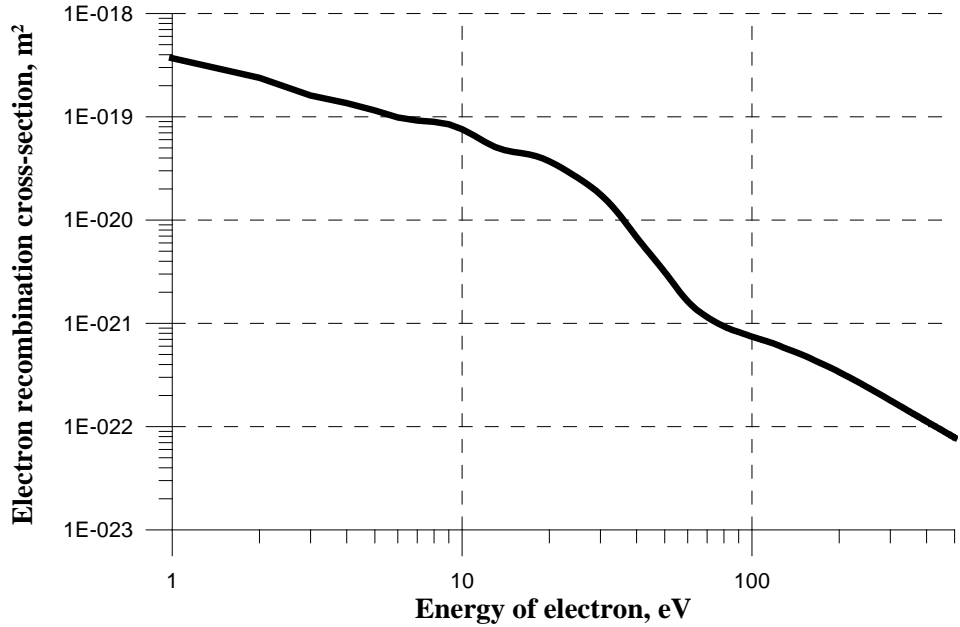

Figure 6: Analytical estimation of cross-sectionof electron capture by fullerene $C_{60}^{+1}$ ion.

Thus, the maximum capture cross-sections are expected at electron energies less about few eV.

We are also interested in caprure cross-sections on ordinary surface-localized electron levels. Because these electron wave functions are localized in thin layer in the vicinity of fullerene sphere surface, then the modules of matrix elements of transition amplitude should be less compared to amplitutudes of transitions on volume-localized electron levels formed by Coulomb field of fullerene ion.

Following the results of the paper [**Ошибка! Закладка не определена.**], the wave functions of delocalized electrons are estimated as

$$\psi_n \approx A \exp\left[-\frac{|r - 6.63|}{1.0}\right], \qquad (20)$$

that is they are exponentially vanish in thin layer in the vicinity of fullerene sphere surface.

The Figure 7 shows VLEL wave function calculated with the use of simplified spherical model ($n=50$) and approximate presentation of SLEL wave function on the basis of the paper [6] after their normalization. One can see that the maximum amplitude of SLEL is larger compared to VLEL if we do not consider a fullerene center where the wave function has the integrable singularity. However, due to the circumstance that the VLEL wavefunction is sinusoidal-like with wave vector from Equation 6, there are the basis for the conclusion that the electron capture cross-section will be larger for VLEL. It is supported by calculations which show that the capture probability on VLELs is 5÷10 times higher in comparison to the capture probability on SLELs.

There is alternative approach to esimate the electron capture cross-section. The general rule connects together the electron recombination (that is the capture) and the photoionization cross-sections using that these phenomena are mutually inverse reactions, see [19] for example:

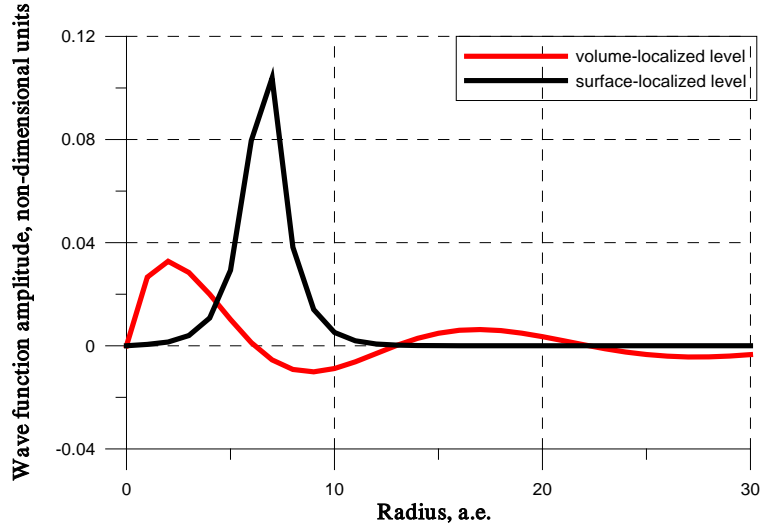

Figure 7: The normalized wave functions for volume-localized electron level with principal quantum number $n$=50 and surface-localized electron level of fullerene.

$$\sigma_{recomb} = \sigma_{photoion} \frac{g_i}{g_k} \frac{2h^2\omega^2}{4\pi^2 m_e^2 c^2 V^2}, \qquad (21)$$

where $\sigma_{recomb}$ is the recombination cross-section, $\sigma_{photoion}$ - the photoionization cross-section, $h$ – the Planck constant, $\omega$ - the angular frequency of photon, $c$ – the speed of light, $m_e$ – the electron mass, $V$ – the electron velocity, $g_i$ and $g_k$ – the statistical weight factors of the photoionization and recombined states respectively. This expression is derived from the detailed partial photoionization cross-section treatment using so called the detailed equilibrium principle. So, the larger the photoionization cross-section the larger the recombination cross-section and inversly.

The analysis of photoionization cross-sections data, see [4-7,20] together with Equation 21 leads us to the conclusion that the electron capture cross-sections as much as $10^{-16} \div 10^{-15}$ cm$^2$ are possible. In the work [21] the recombination cross-sections of the order of $10^{-15}$ cm$^2$ for the electron energy of few eV were obtained from theoretical considerations. These results are in accordance with our predictions.

If the energy of incident electron $E$=100 eV, the electron momentum will be $p \approx$ 0.54·10$^{-23}$ kg·m/s, and its velocity V=$p/m_e \approx$ 5.4·10$^6$ m/s. The electron wave vector is equal to

$$\bar{k} = \bar{p}/\hbar \approx 0.5 \cdot 10^{11} \text{m}^{-1} \qquad (21)$$

Hence, $kr_f \approx 17$. It should be expected (Figure 7), that maximum amplitudes of capture to levels formed by Coulomb field will take place at $kr_f \approx \pi/2$, that is at $k$ one order of magnitude less. For this reason, the maximum of capture cross-section should be at electron energy $E$~1 eV which is observed in Figure 6. The location of this maximum does not depend on ionization extent of fullerene ion.

# The calculation of dipole moments of spontaneous transitions from volume-localized electron levels to other levels of fullerene ions

The dipole electric moment of electron transition from the state $n_1l_1m_1$ to the state $n_2l_2m_2$ is written by the definition as

$$\bar{d}_{n_2l_2m_2,n_1l_1m_1} = e\langle \psi_{n_2l_2m_2} | \bar{r} | \psi_{n_1l_1m_1} \rangle, \qquad (22)$$

Where the wave functions of fullerene ion $C_{60}^{+Z}$ ( Figure 2) within the bounds of simplified spherical model are represented as the product of radial and spherical functions.

The dipole electric moment in Cartesian system of coordinates (we omit the multiplier $e$ – the electron charge) is presented by 3-foldintegral (Figure 2):

$$\bar{d}_{n_2l_2m_2,n_1l_1m_1} = \qquad (23)$$
$$= \int_0^\infty r^2 dr R_{n_1l_1}(r) R_{n_2l_2}(r) \int_0^{2\pi} d\varphi \int_0^\pi d\theta \sin\theta Y_{l_1m_1}(\theta,\varphi) Y_{l_2m_2}(\theta,\varphi) [r\sin\theta\cos\varphi \hat{x} + r\sin\theta\sin\varphi \hat{y} + r\cos\theta \hat{z}],$$

where $\hat{x}$, $\hat{y}$, $\hat{z}$ are the unit vectorsalong correspondingaxes.

The expfressions for dipole electric moment for electron transition from the state $n_1l_1m_1$ to the state $n_2l_2m_2$ includes only $x$- and $z$-components (Figure 2):

$$\bar{d}_{n_2l_2m_2,n_1l_1m_1} = d_x \hat{x} + d_z \hat{z}. \qquad (24)$$

The dipole moments are non-zero for the next transitions:
Let the magnetic quantum number $m_1 = 0$ in initial state $n_1l_1m_1$. Then the x-component of dipole moment $d_x$ is non-zero when $m_1 = \pm 1$. Taking into account normalization of spherical functions (see Appendix 2) it equals to

$$d_x = a(n_1,l_1,n_2,l_2) \cdot \pi K_0(l_1) K_1(l_2) \cdot \int_{-1}^{1} dt P_{l_1}^0(t) \left[ -l_2 t P_{l_2}^0(t) + l_2 P_{l_2-1}^0(t) \right] \qquad (25)$$

where

$$a(n_1,l_1,n_2,l_2) = \int_0^\infty r^3 dr R_{n_1l_1}(r) R_{n_2l_2}(r)$$

$$K_0(l) = \frac{\sqrt{2l+1}}{\sqrt{4\pi}} \qquad (1)$$

$$K_m(l) = \sqrt{\frac{(l-|m|)!}{(l+|m|)!}} \frac{\sqrt{2l+1}}{\sqrt{2\pi}}$$

Let $m_1 = m \neq 0$. Then

$$d_x = a(n_1,l_1,n_2,l_2) \cdot \pi K_m(l_1) K_{m+1}(l_2) \cdot \int_{-1}^{1} dt P_{l_1}^m(t) \left[ -(l_2-m) t P_{l_2}^m(t) + (l_2+m) P_{l_2-1}^m(t) \right] \qquad (27)$$

If $l_1 = l_2 - 1$ the last term in square brackets of (25) and (27) disappears.

At the same time, z-component of dipole moment is non-zero only for $m_1 = m_2 = m$ or for $m_1 = -m_2 = m$ and equals to

$$d_z = a(n_1, l_1, n_2, l_2) \cdot \pi K_m(l_1) K_m(l_2) \cdot \int_{-1}^{1} dt \cdot t \cdot P_{l_1}^m(t) P_{l_2}^m(t) \quad (28)$$

So, for transitions between s-states with azimuthal and magnetic quantum numbers equal to zero, only z-components will be non-zero and from (28) we get:

$$\left| \overline{d}_{n_2 l_2 m_2, n_1 l_1 m_1} \right| = d_z = \frac{1}{4} \int_0^\infty r^3 dr R_{n_1 0}(r) R_{n_2 0}(r) \quad (29)$$

But these transitions between *s*-states of VLELs are excluded due to selection rules. For this reason it is necessary to take into account either transitions from s-states ofVLEL on SLEL with azimuthal *l*=1 or transitions between *p*-states of VLELs (see Appendix 1) and *s*-states of VLELs, see Tables below.

The probability of spontaneous radiation per unit time at dipole transition is equal to

$$P_{n_2 l_2 m_2, n_1 l_1 m_1} = \frac{2\omega^3}{3\hbar c^3 4\pi\varepsilon_0} \left| \overline{d}_{n_2 l_2 m_2, n_1 l_1 m_1} \right|^2 \quad (30)$$

where

$$\hbar\omega = E_{n_1 l_1 m_1} - E_{n_2 l_2 m_2}, \quad \left| \overline{d}_{n_2 l_2 m_2, n_1 l_1 m_1} \right|^2 = d_x^2 + d_z^2 \quad (31)$$

At transition to SI system the calculated dipole moments should be multiplied by

$$\frac{4\pi\varepsilon_0 \hbar^2}{m_e e}$$

So, the probability of spontaneous radiation per unit time and the characteristic time of spontaneous radiation are

$$P_{n_2 l_2 m_2, n_1 l_1 m_1} = \frac{8\pi\omega^3 \hbar^3 \varepsilon_0}{3 c^3 m_e^2 e^2} \left| \overline{d}_{n_2 l_2 m_2, n_1 l_1 m_1} \right|^2$$

$$\tau = \frac{1}{P_{n_2 l_2 m_2, n_1 l_1 m_1}}$$

where the dipole moments are expressed in atomic units as previously.

Due to big size of fullerene the dipole moments of transitions will be considerably larger and characteristic lifetimes of electron states will considerably less in comparison to transitions in ordinary molecules. Also, one can make the assumption analogous to calculation of probability of electron capture in corresponding subsection that the probability of recombination from VLEL to lower VLEL may be higher compared to probability of transition VLEL-SLEL. However, theis prediction is not supported by numerical calculations.

The dipole moments of spontaneous transitions calculated using the formula (22) for fullerene ions between VLELs *Is1*p и *J*s (*J*<*I*) where *I*=2,3,4… and *J*=1,2,3… for *Z*=+1, *Z*=+3, *Z*=+10 are given in Table 1, Table 2 and Table 3 respectively. The three parameters: *x*- component of the dipole moment $\bar{d}_{n_2 l_2 m_2, n_1 l_1 m_1}$ [a.u.], the photon angular frequency $\omega$ [s$^{-1}$] and the characteristic lifetime of the level $Ip$ $\tau$ [s] are presented in these Tables.

Table 1: The dipole moments of transitions, the angular frequency and the characteristic $Ip$ level's lifetime for the fullerene ion $C_{60}^{+1}$

| Js \ Ip | 2p | 3p | 5p | 10p | 50p |
|---|---|---|---|---|---|
| SLEL | -0.24<br>5.6·10$^{15}$<br>3.2·10$^{-7}$ | 0.80<br>8.5·10$^{15}$<br>8.4·10$^{-9}$ | 0.66<br>1.0·10$^{16}$<br>7.8·10$^{-9}$ | 0.44<br>1.1·10$^{16}$<br>1.5·10$^{-8}$ | 0.20<br>1.1·10$^{16}$<br>6.5·10$^{-8}$ |
| 2s | – | 0.57<br>2.9·10$^{15}$<br>4.3·10$^{-7}$ | 0.64<br>4.3·10$^{15}$<br>9.8·10$^{-8}$ | 0.48<br>5.0·10$^{15}$<br>1.2·10$^{-7}$ | 0.26<br>5.2·10$^{15}$<br>3.5·10$^{-7}$ |
| 3s | – | – | -1.49<br>1.5·10$^{15}$<br>4.7·10$^{-7}$ | -1.45<br>2.1·10$^{15}$<br>1.7·10$^{-7}$ | -1.32<br>2.3·10$^{15}$<br>1.6·10$^{-7}$ |
| 5s | – | – | – | -0.80<br>6.2·10$^{14}$<br>2.2·10$^{-5}$ | -4.34<br>8.2·10$^{14}$<br>3.2·10$^{-7}$ |
| 10s | – | – | – | – | 2.17<br>2.0·10$^{14}$<br>9.0·10$^{-5}$ |

Table 2: The dipole moments of transitions, the angular frequency and the characteristic $Ip$ level's lifetime for the fullerene ion $C_{60}^{+3}$

| Js \ Ip | 4p | 6p | 9p | 15p | 45p |
|---|---|---|---|---|---|
| SLEL | 0.98<br>1.2·10$^{16}$<br>2.2·10$^{-9}$ | 0.75<br>1.8·10$^{16}$<br>9.9·10$^{-10}$ | 1.08<br>2.1·10$^{16}$<br>3.1·10$^{-10}$ | 1.89<br>2.2·10$^{16}$<br>8.2·10$^{-11}$ | 0.99<br>2.3·10$^{16}$<br>2.7·10$^{-10}$ |
| 4s | – | -1.04<br>6.5·10$^{15}$<br>1.1·10$^{-8}$ | -0.76<br>9.3·10$^{15}$<br>7.0·10$^{-9}$ | -1.45<br>1.1·10$^{16}$<br>1.2·10$^{-9}$ | -0.98<br>1.2·10$^{16}$<br>2.3·10$^{-9}$ |
| 6s | – | – | 0.93<br>2.9·10$^{15}$<br>1.6·10$^{-7}$ | 1.76<br>4.3·10$^{15}$<br>1.3·10$^{-8}$ | 1.0<br>5.1·10$^{15}$<br>2.5·10$^{-8}$ |
| 9s | – | – | – | 1.81<br>1.5·10$^{15}$<br>3.2·10$^{-7}$ | 0.78<br>2.2·10$^{15}$<br>5.1·10$^{-7}$ |
| 15s | – | – | – | – | 0.41<br>7.3·10$^{14}$<br>5.1·10$^{-5}$ |

Table 3: The dipole moments of transitions, the angular frequency and the characteristic $Ip$ level's lifetime for the fullerene ion $C_{60}^{+10}$

| Js \ Ip | 6p | 10p | 30p | 50p | 100p |
|---|---|---|---|---|---|
| SLEL | 0.24<br>9.5·10$^{15}$<br>6.8·10$^{-8}$ | -2.43<br>4.6·10$^{16}$<br>5.7·10$^{-12}$ | 4.22<br>6.5·10$^{16}$<br>6.9·10$^{-13}$ | 4.38<br>6.6·10$^{16}$<br>6.0·10$^{-13}$ | 4.11<br>6.7·10$^{16}$<br>6.6·10$^{-13}$ |
| 6s | – | 0.30<br>3.7·10$^{16}$<br>7.4·10$^{-10}$ | 1.36<br>5.5·10$^{16}$<br>1.1·10$^{-11}$ | 1.04<br>5.7·10$^{16}$<br>1.7·10$^{-11}$ | 0.76<br>5.7·10$^{16}$<br>3.0·10$^{-11}$ |
| 10s | – | – | -1.94<br>1.8·10$^{16}$<br>1.4·10$^{-10}$ | -2.31<br>2.0·10$^{16}$<br>7.9·10$^{-11}$ | -2.38<br>2.1·10$^{16}$<br>6.8·10$^{-11}$ |
| 30s | – | – | – | -3.73<br>1.5·10$^{15}$<br>7.5·10$^{-8}$ | -3.64<br>2.1·10$^{15}$<br>2.7·10$^{-8}$ |
| 50s | – | – | – | – | -2.58<br>6.2·10$^{14}$<br>2.1·10$^{-6}$ |

From the presented Tables one can see that the calculated dipole moments depend on fullerene ionization extent, initial and final electron state, and are varied in broad range from about 0.2 to 5 in atomic system of units. Basically, the dipole moments of transition VLEL-SLEL are of the same order of magnitude as the dipole moment of transition VLEL-VLEL.

**The analysis of possibility of coherent radiation generation on fullerene C$_{60}$ ions VLELs**

The analysis conducted in previous Sections gives ground to use the extraordinary properties of fullerene ions for coherent radiation generation.

There is a principal possibility to get the coherent radiation using the medium containing two parts of particles: excited fullerene $C_{60}^{+Z}$ ions having with one electron at VLEL state $n_1 l_1 m_1$ (state 1) and excited fullerene $C_{60}^{+Z}$ ions with one electron at VLEL state $n_2 l_2 m_2$ (state 2) or fullerene $C_{60}^{+Z}$ ions in equilibrium state. The quantum energy level 1 is higher than the level 2.

Let us introduce the following designations: $n_1$ is the concentration of excited fullerene ions with electrons at the state 1, $\omega$ and $\lambda$ - the radiation frequency and corresponding wavelength at transition $1 \rightarrow 2$,

To estimate the inverse concentration $n_1$ needed for attainment of generation threshold for coherent radiation at VLELs we cah use a simple formulation:

$$\mu_\omega L_{abs} \gg 1$$

where

$$\mu_\omega = \frac{\lambda^2}{2\pi} n_1 \frac{\Delta\omega}{\Delta\omega_{sp}}$$

Here $\mu_\omega$ is the coefficient of resonance amplification per unit length;

$L_{abs}$ - the absorption length of photons;

$\Delta\omega_{sp}$ - the width of dipole spontaneous radiation line.

$\Delta\omega$ - the total broadening of emission line due to Doppler effect, collisions, radiationless losses and spontaneous radiation: $\Delta\omega = \Delta\omega_{dop} + \Delta\omega_{col} + \Delta\omega_{nr} + \Delta\omega_{sp}$.

To reach the generation threshold it is necessary to have

$$n_1 \gg \frac{2\pi}{\lambda^2 L_{abs}} \frac{\Delta\omega_{sp}}{\Delta\omega}$$

The absorption spectrum of fullerene gas $C_{60}$ was investigated in the paper [22]. According to experimentally obtained data given in this paper the absorption cross-section is about $\sigma_{abs} \sim (1 \div 5) \cdot 10^{-15} \text{cm}^2$ in the wavelength range $\lambda = 200 \div 400$ nm.

The length of absorption is equal to

$$L_{abs} \approx \frac{1}{n_{full} \sigma_{abs}}$$

where the concentration of neutral fullerenes $n_{full} \sim 10^{17} \text{cm}^{-3} = 10^{23} \text{m}^{-3}$ for fullerene gas at the temperatute $T \cong 700°C$.

Basically, for the estimation we may put $\Delta\omega_{sp} \sim \Delta\omega$. If we use the characteristic value for the wavelength $\lambda \sim 10^{-9} \div 10^{-7}$ m which corresponds to $\omega \sim 5 \cdot 10^{14} \div 5 \cdot 10^{16}$ 1/s (see Tables 1–3) then the generation threshold for inverse level concentration will be in the range

$$n_1 \gg 6 \cdot 10^8 \div 6 \cdot 10^{12} \text{cm}^{-3}$$

which is possible because the concentration of fullerene ions $C_{60}^{+Z}$ under the electron beam interaction conditions diminishes by about one order of magnitude with growth of ionization degree $Z$ by unity and the electron capture cross-sections are large enough.

For example, if the gas with concentration of neutral fullerenes about $n_{full} = 10^{17} \text{cm}^{-3}$ then, according to results of electron capture calculations, it is possible to get the concentration of $C_{60}^{+Z}$ ions about

$$n_f^{+3} \approx 10^{14} \text{cm}^{-3}.$$

Thus, the estimations show that the medium containing fullerene $C_{60}^{+Z}$ ions may be used for the generation of coherent radiation.

**Conclusions**

The existence of volume-localized electron levels (VLELs) of fullerene ions $C_{60}^{+Z}$ is proven on the basis of simplified quantum-mechanical consideration of fullerene ion. These levels arise in Coulomb potential well formed inside fullerene after its ionization. The simplified calculation of VLELs wave functions is conducted using the fullerene sphere model. The basic parameters of VLELs including energy and spatial functional dependencies are obtained. To get those characteristics the analytical dependencies of wave functions inside and outside the fullerene sphere are used which are sewn on the sphere. We use the model proposition that the charge is uniformly distributed on the sphere that is the simplified spherical model approach.

The analysis of electron capture during the interaction of electron beam with the medium consistent of fullerenes ions $C_{60}^{+Z}$ is conducted as the application of the methods developed.

The calculation estimations done show that the electron capture cross-sections for capture on VLELs are about one order of magnitude larger compared to capture cross-sections on ordinary SLELs at electron energy of the order 10 eV and is about $10^{-19}$ m$^2$.

The calculations of dipole moments for transitions from fullerene ions VLELs to other VLELs and to SLELs are also conducted. The calculated dipole moments depend on fullerene ionization extent, initial and final electron state, and are varied from about 0.2 to 5 in atomic system of units.

The unique features of fullerenes give ground to new interesting opportunities. The principal possibility of coherent radiation generation on fullerene ions VLELs is discussed.

**Acknowledgments**

# Appendix 1. Expressions for radial functions with azimuthal number *l*>0 and the presentation of wave functions outside fullerene sphere

Consider the quantum states with $l>0$, $l \leq n-1$. To determine corresponding wave functions it is convenient to use the following recurrent expression:

$$R_{nl} = r^l \chi_{nl}$$

$$\chi_{n,l+1} = \frac{1}{r}\frac{d\chi_{nl}}{dr}$$

From which it is easy to get all $R_{nl}$ in consecutive order once you know $R_{1l}$.

For example, inside fullerene (*r*< 6.63) we have:

$$R_{n0} = \frac{\sin[k_n r]}{r},$$

$$R_{n1} = -\frac{\sin[k_n r]}{r^2} + \frac{k_n}{r}\cos[k_n r], \qquad (32)$$

$$R_{n2} = \frac{\sin[k_n r]}{r^3}[3 - k_n^2 r^2] - \frac{3k_n}{r^2}\cos[k_n r],$$

$$R_{n3} = \frac{\sin[k_n r]}{r^4}[6k_n^2 r^2 - 15] + \frac{k_n}{r^3}[15 - k_n^2 r^2]\cos[k_n r], \; etc.$$

Here

$$k_n = \sqrt{2\left(E_n + \frac{Z}{r_f}\right)}.$$

Analogous expression can be consecutively obtained if $\cos[k_n r]/r$ is given as $R_{n0}$.

The analysis of expressions (32) shows that all $R_{nl}$ satisfy the conditions of finiteness of integral from wave function module square. This is why asimptotics of functions $R_{nl}$ at $r \to 0$ is as following:

$$R_{nl} \approx \frac{2k_n^{l+1}}{(2l+1)!!} r^l$$

Then wave functions inside and outside fullerene sphere should be sewn to ensure continuity of wave function and its first derivative (it is practically useful to equate logarithmic derivative of wave function inside and outside sphere), see Figure 8.

This discussion leads us to the conclusion that the wave eigenfunctions of VLELs are the expressions (8) for azimuthal and magnetic quantum numbers $l$ and $m$ equal to 0. For $l > 0$ the expressions $R_{n1}$ from (32) serve as radial parts of VLELs because they have integrated singularity at $r=0$.

We use the following quantity as an energy unit

$$\frac{m_e e^4}{\hbar^2 (4\pi\varepsilon_0)^2} \approx 27.21 \text{ эВ.} \tag{33}$$

In any case the spatial dependencies of wave functions with $l=0$ have the maximum in the center of fullerene spheroid as it is shown in Figure 8.

Consider in more detail the solutions for wave functions outside the fullerene sphere, see Equation 4. Rewrite this equation for the case $l=0$:

$$\frac{d^2 R_{n0}}{dr^2} + \frac{2}{r}\frac{dR_{n0}}{dr} + \frac{2Z}{r} R_{n0} = -2E_{n0} R_{n0} \tag{34}$$

or

$$\frac{d}{dr}\left(r^2 \frac{dR_{n0}}{dr}\right) + 2Zr R_{n0} = -2E_{n0} r^2 R_{n0} \tag{35}$$

Realizing the following substitute

$$u_{n0} = R_{n0} r \tag{36}$$

we get

$$\frac{d^2 u_{n0}}{dr^2} + 2Z \frac{u_{n0}}{r} = -2E_{n0} u_{n0}, \tag{37}$$

After substitution $\hat{r} = 2r$ this equation transforms to the following view:

$$\frac{d^2 u_{n0}}{d\hat{r}^2} + Z \frac{u_{n0}}{\hat{r}} = -\frac{E_{n0}}{2} u_{n0}. \tag{38}$$

In the case of inequality $-rE_{n0} \gg Z$ that is at rather large distances from the centre the following equation will be valid:

$$\frac{d^2 u_{n0}}{d\hat{r}^2} = -\frac{E_{n0}}{2} u_{n0}, \quad (39)$$

which has the next solutions:

$$u_{n0} = e^{-\hat{r}\sqrt{-\frac{E_{n0}}{2}}}, \quad u_{n0} = e^{+\hat{r}\sqrt{-\frac{E_{n0}}{2}}}, \quad (40)$$

The second solution is excluded, because it tends to infinity at large distances.
We look now for the solution in the form

$$u_{n0} = f_{n0}(\hat{r}) e^{-\hat{r}\sqrt{-\frac{E_{n0}}{2}}} \quad (41)$$

Then for $f_{n0}(\hat{r})$ it is valid

$$\frac{d^2 f_{n0}}{d\hat{r}^2} - 2\sqrt{-\frac{E_{n0}}{2}} \frac{df_{n0}}{d\hat{r}} + \frac{Z f_{n0}}{\hat{r}} = 0 \quad (42)$$

After next substitute

$$\hat{r} = \frac{1}{\sqrt{-2E_{n0}}} x \quad (43)$$

we get

$$\frac{d^2 f_{n0}}{dx^2} - \frac{df_{n0}}{dx} + \frac{Z}{\sqrt{-2E_{n0}}} \frac{f_{n0}}{x} = 0 \quad (44)$$

Let us designate

$$\xi_{n0} = \frac{Z}{\sqrt{-2E_{n0}}} \quad (45)$$

Then

$$\frac{d^2 f_{n0}}{dx^2} - \frac{df_{n0}}{dx} + \xi_{n0} \frac{f_{n0}}{x} = 0 \quad (46)$$

The Equation 46 has the solution limited at $x \to \infty$ only for integer $\zeta_{n0}=1,2,3...$ For the solution of this equation it is convenient to search the solution in the form of infinite set:

$$f_{n0}(x) = \sum_{i=n}^{\infty} a_i x^i \quad (47)$$

Substituting (43) into (42) we get:

$$\sum_{i=n}^{\infty} \left[ i(i-1) a_i x^{i-2} - i a_i x^{i-1} + \xi_{n0} a_i x^{i-1} \right] = 0. \quad (48)$$

By substituting of summation index $i = j + 1$ we represent the sum on first term in square brackets this way:

$$\sum_{i=n}^{\infty} i(i-1) a_i x^{i-2} = \sum_{j=n-1}^{\infty} j(j+1) a_{j+1} x^{j-1}. \quad (49)$$

Hence,

$$n(n-1) a_n + \sum_{i=n}^{\infty} \left[ i(i+1) a_{i+1} x^{i-1} - i a_i x^{i-1} + \xi_{n0} a_i x^{i-1} \right] = 0. \quad (50)$$

Let the principal quantum number be $n=1$. We can choose $a_1 = 1$. Then we get the following recurrence formula for determination of coefficients $a_i$:

$$a_1 = 1; \qquad a_{i+1} = \frac{(i - \xi_{10})}{i(i+1)} a_i \qquad (51)$$

that is

$$a_1 = 1; \qquad a_i = \frac{(i-1-\xi_{10}) \cdot (i-2-\xi_{10}) \cdot \ldots \cdot (2-\xi_{10}) \cdot (1-\xi_{10})}{n \cdot (n-1)! \cdot (n-1)!} \qquad (52)$$

Let the principal quantum number now be $n > 1$. Then

$$a_i = 0, \qquad i \leq n; \qquad (53)$$

$$a_{n+1} = 1; \qquad a_{i+1} = \frac{(i - \xi_{10})}{i(i+1)} a_i, \qquad i > n \qquad (54)$$

If the azimuthal quantum number $l$ is more than 0, then the radial part of wave function satisfies to the following equation:

$$\frac{d^2 f_{nl}}{dx^2} - \frac{df_{nl}}{dx} + \zeta_{nl} \frac{f_{nl}}{x} - \zeta_{nl}^2 l(l+1) \frac{f_{nl}}{x^2} = 0. \qquad (55)$$

Look for solution in the form

$$f_{nl}(x) = \sum_{i=n}^{\infty} a_i x^i \quad n=l+1, l+2, \ldots \qquad (56)$$

$$a_1 = 0, \ a_2 = 0, \ a_3 = 1, \ a_{i+1} = \frac{(i - \xi_{nl})}{i(i+1) - \xi_{nl}^2 l(l+1)} a_i.$$

For radial functions $R_{nl}(r)$ the following condition of normalization is valid:

$$\int_0^\infty r^2 dr R_{n_1 l_1}(r) R_{n_2 l_2}(r) = 0 \qquad n_1 \neq n_2 \qquad \forall l_1, l_2,$$

$$\int_0^\infty r^2 dr R_{n_1 l_1}(r) R_{n_2 l_2}(r) = \delta_{l_1 l_2} \qquad n_1 = n_2,$$

where $\delta_{ik}$ is a symbol of Kronecker.

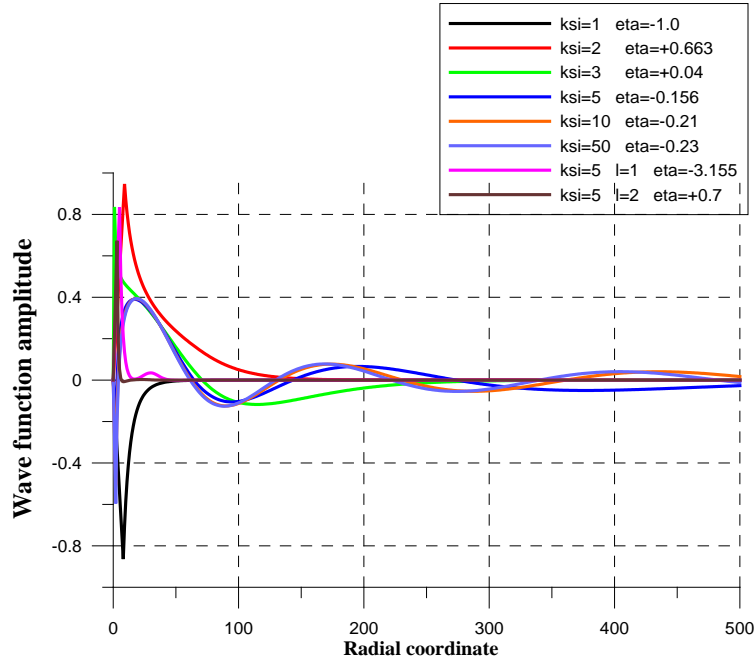

Figure 8: The wave functions of VLELsof fullerene $C_{60}^{+1}$ ion with different parameters $\zeta$ (principal quantum number), $l$ (azimuthal quantum number) and $\eta$ (the "sewing" parameter).

# Appendix 2. A summary of basic formulas for normalization of spherical functions

In expression for wave functions with azimuthal number $l$ for fullerene ion within the bounds of simplified spherical model there are $(2l+1)$ independent spherical functions:

$$Y_{lm}(\theta,\varphi) = P_l^m(\cos\theta)\cos m\varphi, \quad m = 0,1...l,$$

$$Y_{lm}(\theta,\varphi) = P_l^m(\cos\theta)\sin|m|\varphi, \quad m = -1,-2...-l,$$

where $P_l^m(\cos\theta)$ are attached functions of Legendre:

$$P_l^m(t) = \left(1-t^2\right)^{\frac{|m|}{2}} \frac{d^{|m|}}{dt^{|m|}} P_l(t).$$

Here

$$P_l(t) = \frac{1}{2^l l!} \frac{d^l}{dt^l}\left(t^2 - 1\right)^l$$

- the Legendre polynomials.
The following recurrence formula takes place for them:

$$(l+1)P_{l+1}(t) - (2l+1)tP_l(t) + lP_{l-1}(t) = 0, -1 \leq t \leq 1, \quad l \geq 0.$$

The normalization of spherical functions is as following:

$$\left(Y_{l_1 m_1}, Y_{l_2 m_2}\right) = \int_0^{2\pi} d\varphi \int_0^{\pi} d\theta \sin\theta Y_{l_1 m_1}(\theta,\varphi) Y_{l_2 m_2}(\theta,\varphi)$$

$$\left(Y_{l_1 m_1}, Y_{l_2 m_2}\right) = \int_0^{2\pi} d\varphi \cos m_1\varphi \cos m_2\varphi \int_0^{\pi} d\theta \sin\theta P_{l_1}^{m_1}(\cos\theta) P_{l_2}^{m_2}(\cos\theta)$$

$$\left(Y_{l_1 m_1}, Y_{l_2 m_2}\right) = \int_0^{2\pi} d\varphi \cos m_1\varphi \cos m_2\varphi \int_{-1}^{1} dt P_{l_1}^{m_1}(t) P_{l_2}^{m_2}(t)$$

$$\left(Y_{l_1 m_1}, Y_{l_2 m_2}\right) = 0 \qquad l_1 \neq l_2 \text{ либо } m_1 \neq m_2$$

$$\left(Y_{lm}, Y_{lm}\right) = \frac{4\pi}{2l+1}, \quad m=0$$

$$\left(Y_{lm}, Y_{lm}\right) = \frac{(l+|m|)!}{(l-|m|)!} \frac{2\pi}{2l+1}, \quad 1 \leq |m| \leq l$$

So, to ensure the normalization of spherical function $Y_{lm}(\theta,\varphi)$, it is necessary to multiply it by a factor

$$\frac{\sqrt{2l+1}}{\sqrt{4\pi}}, \qquad m=0,$$

$$\sqrt{\frac{(l-|m|)!}{(l+|m|)!}} \frac{\sqrt{2l+1}}{\sqrt{2\pi}}, \quad 1 \leq |m| \leq l.$$